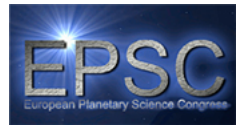

# Contribution of amateur observations to Saturn storm studies

**M. Delcroix** (1) and G. Fischer (2)
(1) Commission des observations planétaires, Société Astronomique de France (delcroix.marc@free.fr) (2) Space Research Institute, Austrian Academy of Sciences (georg.fischer@oeaw.ac.at)

## Abstract

Since 2004, Saturn Electrostatic Discharges (SEDs), which are the radio signatures of lightning in Saturn's atmosphere, have been observed by the Cassini Radio and Plasma Wave Science instrument (RPWS). Despite their important time coverage, these observations lack the resolution and positioning given by imaging around visible wavelengths. Amateur observations from Earth have been increasing in quality and coverage since a few years, bringing information on positions, drift rates and shape evolutions of large visible white spots in Saturn's atmosphere. Combining these two complementary sources has brought better analysis of Saturn's storms evolutions.

## 1. Introduction

The Cassini Imaging Science Subsystem (ISS), and space or earth based professional telescopes have many observational targets and can only provide a limited coverage of Saturn's atmospheric features. Here many amateur astronomers can provide a more extended coverage, which proved to be useful in lightning storms studies.

## 2. Storm observations

### 2.1 Amateur observations

Instruments used by amateurs are mostly reflectors with an aperture from 15 to 40 cm. Since 2001, image coverage during Saturn's apparition has increased drastically: The spot positions from images of more than 180 observers have been measured yielding 780 large white spots positions detected in Saturn's atmosphere at all latitudes.

The software used for these measurements, winjupos, has been used by amateur astronomer associations since 2003 to measure positions, calculate drifts and produce reports on features evolutions on Jupiter.

The number of positional measurements of long lived white spots (called "storms") visible in the South Tropical Zone (nicknamed "storm alley") is summarized in Table 1 for each apparition.

Table 1: Storm measurements (2004-2009). A1 and A2 denote different longitude ranges, so do B1 and B2, and C1, C2, and C3 (see Figure 1 for the latter).

| Apparition-Storm reference | No. of positions | Time range |
| --- | --- | --- |
| 2003-2004-A1 | 12 | 13/09/03-01/02/04 |
| 2003-2004-A2 | 9 | 12/12/03-27/03/04 |
| 2005-2006-B1 | 9 | 22/01/06-20/04/06 |
| 2005-2006-B2 | 11 | 24/01/06-25/02/06 |
| 2007-2008-C1 | 23 | 01/12/07-18/06/08 |
| 2007-2008-C2 | 13 | 01/12/07-23/03/08 |
| 2007-2008-C3 | 20 | 14/03/08-18/06/08 |
| 2008-2009-D | 18 | 07/12/08-11/06/09 |

These measurements were done mainly with color/integrated light images, sometimes red (high-pass or band-pass) filtered images and rarely green or blue band-pass filtered images.

### 2.2 Cassini radio observations

The information which can be correlated with amateur observations is the occurrence in time, the number of SEDs, and the estimated longitudinal drift of the SED episodes.

## 3. Results

### 3.1 Amateur observations

From amateur data, latitude position and longitudinal drift rates have been computed (cf. Table 2). White spots aligned along the same drift line were considered as the same "storm" (cf. graphs such as Figure 1, in [1] and [2] for 2008-2009 storm).

Table 2: Latitude/drift rates of storms (2004-2009)

| Apparition-Storm reference | Latitude (° centric) | LIII drift rate (°/day) |
| --- | --- | --- |
| 2003-2004-A1 | -35.3 +/-0.2 | 0.289 +/-0.020° |
| 2003-2004-A2 | -35.2 +/-0.2 | 0.265 +/-0.015° |
| 2005-2006-B1 | -35.1 +/-0.4 | 1.033 +/-0.021 |
| 2005-2006-B2 | -34.9 +/-0.2 | 0.597 +/-0.035° |
| 2007-2008-C1 | -35.0 +/-0.2 | 0.337 +/-0.009° |
| 2007-2008-C2 | -34.4 +/-0.4 | 0.983 +/-0.014° |
| 2007-2008-C3 | -34.9 +/-0.2 | 0.354 +/-0.026° |
| 2008-2009-D | -35.0 +/-0.2 | 0.303 +/-0.007° |

The shapes of the storms can only be distinguished on the best images. As observations come from non calibrated sources, it is not possible to derive luminosity figures, but the major brightness changes can be followed.

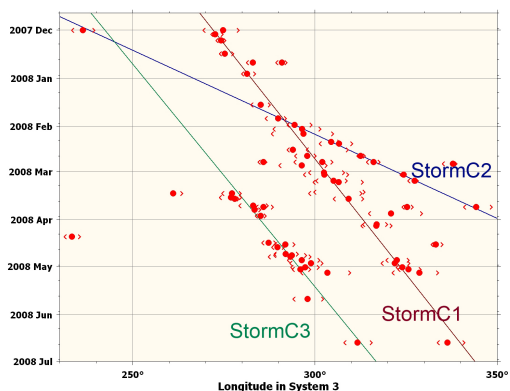

Figure 1: Storms and drifts in 2007-2008 apparition

### 3.2 Saturn Electrostatic Discharges

From the longitude range of SED episodes, and taking into account that the radio horizon is larger than the visible horizon, approximate positions and drifts of the storms have been estimated (see [3] and [4]). These estimations were in agreement with Cassini ISS observations (see [5]) that showed convective cloud features at a planetocentric latitude of -35°.

## 4. Comparison of observations from amateurs and Cassini RPWS

"Storms" observed by amateurs are as well located around a latitude of -35°. The drifts are in good agreement for 2003-2004 (0.265°/d vs. 0.32°/d - see [3]), for 2006 B1 (0.597°/d vs. 0.6°/d – see [4]), and also for observations between 2007 (0.337°/d vs. 0.3°/d for 2007-2008 C1) and 2009. The longitudes of the storms are coherent in all observations.

Whenever SED activity is strong, amateur storm observations are possible and bright, showing inertia for brightness changes related to SED activity of a few days. In March 2008 SEDs have been observed without interruptions for several Saturn rotations which cannot be explained by one single source. This was coherent with several storms being visible in amateur images at the same time (see Figure 1), proving lightning occurred in more than one storm at the same time.

## 5. Summary and Conclusions

Useful information about Saturn's storms is derived from amateur astronomers' images: drift rates, brightness changes, and simultaneous features observations. This proved to be coherent with SEDs detections since 2004, allowing regular observations in visible wavelengths of storms that are associated with lightning detected at radio frequencies, giving a complementary coverage of such complex features.

## Acknowledgements

We would like to thank all amateurs who have been providing their images for their time and dedication.

## References

[1] Delcroix, M.: Opposition de Saturne 2006-2007, Observations & Travaux, n°72, pp 2-16, 2009.

[2] Delcroix, M.: La plus longue tempête sur Saturne, L'Astronomie, n° 21, pp44-45, 2009.

[3] Fischer, G. et al.: Saturn lightning recorded by Cassini/RPWS in 2004, Icarus 183, pp135-152, 2006.

[4] Fischer, G. et al.: Analysis of a giant lightning storm on Saturn, Icarus 190, pp528-544, 2007.

[5] Dyudina, U. et al.: Lightning storms on Saturn observed by Cassini ISS and RPWS during 2004-2006, Icarus 183, pp545-555, 2007.